\documentclass[acmsmall]{acmart}
\usepackage{booktabs}

\AtBeginDocument{
  }

\setcopyright{cc}
\setcctype{by}
\acmJournal{PACMHCI}
\acmYear{2026} \acmVolume{10} \acmNumber{6} \acmArticle{CSCW068}
\acmMonth{10} \acmDOI{10.1145/3816916}

\begin{document}

\title[Ctrl-F-Resist]{Ctrl-F-Resist. Practices, Challenges, and Technical Needs of Civil Society Organizations Monitoring the Far-Right Online}

\author{Elisabeth Steffen}
\email{steffen@htw-berlin.de}
\orcid{0000-0003-0170-968X}
\affiliation{
  \institution{Hochschule f\"ur Technik und Wirtschaft, University of Applied Science, Berlin}  \country{Germany}
}
\affiliation{
  \institution{Technische Universit\"at, Technical University, Berlin}
    \country{Germany}
}

\author{Helena Mihaljevi\'c}
\email{helena.mihaljevic@htw-berlin.de}
\orcid{0000-0003-0782-5382}
\affiliation{
  \institution{Hochschule f\"ur Technik und Wirtschaft, University of Applied Science, Berlin}    \country{Germany}
}

\begin{abstract}

As far-right actors increasingly exploit online platforms to disseminate ideology and mobilize supporters, civil society organizations (CSOs) play a vital yet under-recognized role in monitoring antidemocratic dynamics online. Unlike fact-checkers or content moderators, CSOs engage in long-term, contextualized analysis, often in resource-constrained settings and under precarious working conditions. Despite their critical societal role, CSOs face significant barriers to adopting or co-developing technical solutions, including legal uncertainty, limited platform access, and chronic underfunding. Existing research and tool development efforts have largely overlooked these actors in favor of more institutionally embedded stakeholders.

This paper addresses this gap through a qualitative study with 15 practitioners from 12 Germany-based CSOs engaged in online monitoring, positioning them as key yet overlooked stakeholders in the governance of digital spaces. We explore their current practices, challenges, and expectations regarding technological support. Our findings show that monitoring remains largely manual due to the lack of tailored tools, with enhanced search capabilities emerging as the most pressing technical need. While participants express openness to AI-supported features such as media processing and content discovery, many remain skeptical of automated classification, citing concerns around trust, legal usability, and professional credibility.

Grounded in these findings, we introduce a conceptual monitoring workflow and describe its implementation in an open-source Telegram monitoring prototype designed to flexibly support diverse monitoring goals. To account for the sociotechnical character of CSO monitoring, we outline concrete design recommendations as well as recommendations for policymakers and academic research. In addition, we introduce the manual labor trap as an empirically grounded concept that explains why monitoring CSOs tend to remain locked into labor-intensive, low-capacity arrangements, and propose to develop a CSCW-grounded theory of collaborative sociotechnical scaling to better understand both the promise and challenges of collaboration as a pathway beyond fragmented, labor-intensive monitoring work.

\end{abstract}



\maketitle

\section{Introduction}
Over the past decades, antidemocratic behavior in online spaces has become increasingly visible, normalized and strategically organized \cite{eu_parliament_2024,adl_hate_report_2023}. Far-right actors have developed considerable media literacy \cite{conway_right-wing_2019}, strategically utilizing discussion forums \cite{aakerlund_2021_influence}, mainstream platforms like TikTok \cite{widholm_etal_2024,classen_2024_afd} and Instagram \cite{mellea_2024_activeclubs,beauduin_2025_socialmedia}, as well as `alternative', largely unmoderated spaces like Telegram \cite{rothut_etal_2024,guhl_etal_2020,schulze_2020} to mobilize supporters and disseminate ideology. These developments intensify during crises and significant policy changes, e.\,g. during the COVID-19 pandemic or in the context of elections \cite{isd_afd_2024}. In the context of these events, right-wing and conspiratorial groups amplified their narratives and mobilization efforts via online platforms \cite{darius_2022_hashjacking,buehling_heft_2023,bbc_2024_farmersprotest} which function not only as spaces for coordination and mobilization but also as arenas for performative self-promotion, including live video streams of demonstrations \cite{democ_interferenzen_2023}, visual documentations of criminal acts \cite{karimi_cnn_2021} or violent attacks against individual political opponents \cite{tagesspiegelGewaltErpressung}.
\textbf{Monitoring} of respective dynamics is key to understanding and mitigating potential harms for individuals and society.

Computational research is increasingly addressing the needs of practical stakeholders, especially by developing algorithmic methods and tools to detect, analyze, or mitigate online harm \cite{fortuna_survey_2018,yang_multimodal_2022,guo_survey_2022}. However, much of this research has focused on fact-checking (FC) and content moderation (CM), both in application scenarios and in the design of user studies, while the specific needs, constraints, and workflows of \textbf{civil society organizations (CSOs)}, remain underexplored \cite{baumler_harnessing_2025}.  In particular, the design of technical tools tailored to CSOs' monitoring requirements remains an open challenge.

To address this gap, this paper examines how CSOs monitor and analyze antidemocratic dynamics on social media, based on a qualitative study with 15 practitioners from 12 Germany-based organizations engaged in online monitoring. Our contribution is threefold: first, we provide an in-depth empirical account of CSO monitoring practices under complex organizational and political conditions; second, we derive design implications and integrate them into an ongoing co-creation process to develop a durable open-source tool; and third, we introduce the manual labor trap as an empirically grounded concept and propose to develop a CSCW-grounded theory of collaborative sociotechnical scaling that helps to understand how collaborative infrastructures might enable alternative scaling trajectories.

Our investigation is guided by the following three research questions:

\begin{itemize}
\item RQ1: What are the \textbf{current practices and workflows} employed by civil society organizations to monitor social media for antidemocratic content and actors?
\item RQ2: What \textbf{technical, ethical, and legal challenges} do these organizations encounter in their monitoring efforts?
\item RQ3: What are their \textbf{needs and expectations regarding technological support}?
\end{itemize}

Our findings show that online monitoring by CSOs remains largely manual, primarily due to the lack of tailored technical tools. Persistent ethical and regulatory challenges hinder cross-organizational collaboration -- both in terms of data sharing and coordination of content-related efforts -- which, in turn, limit the potential for resource-efficient, collaborative technical innovation. 

The most frequently articulated technical need is for improved search capabilities across a broad range of content types and modalities, including audio, images, and video.  AI-based automation is perceived as potentially useful for media processing tasks such as transcription or optical character recognition (OCR). However, participants expressed substantial skepticism toward automated content classification, highlighting a gap between current research trends and the practical realities faced by CSOs. Trust plays a central role: AI is often viewed not only as unreliable, but also as a potential threat to legal validity, professional credibility, and the value of human expertise.

We introduce a conceptual monitoring workflow grounded in our empirical findings and demonstrate its implementation in a monitoring software prototype through eight new components which are designed to flexibly support the diverse thematic interests of monitoring CSOs and to be configurable through lightweight default settings that accommodate their varying and often limited technical capacities.

This study contributes to the CSCW community by foregrounding civil society actors as key yet often overlooked stakeholders in the governance of digital spaces, and by showing how grassroots monitoring operates under precarious, adversarial, and politically charged conditions, reflecting CSCW research on invisible work and articulation work \cite{star_layers_1999,schmidt_taking_2011}. Moreover, this work contributes to HCI research on human-centered AI \cite{auernhammer_human-centered_2020,bingley_where_2023} within the context of civil society online monitoring. 
Our empirical and design-oriented insights identify what is needed to enable sustainable civic technologies that align with the values and working conditions of those on the front lines of defending democracy online. 
We provide six directions for future CSCW research that could deepen engagement with civil society needs, advance responsible technology integration, and foster more inclusive forms of digital governance. We further make three recommendations for policy makers and ten recommendations for the design and development of monitoring tools. 
Finally, we extend CSCW scholarship on sociotechnical infrastructuring \cite{star_steps_1994,neumann_making_1996,karasti_infrastructuring_2014} and heterogeneous, collaborative knowledge production \cite{ackerman_sharing_2013,star_institutional_1989} by introducing the empirically grounded concept of the manual labor trap, and suggest the development theory of collaborative sociotechnical scaling, which positions collaboration as a pathway beyond fragmented, labor-intensive workflows while also foregrounding the challenges inherent in inter-organizational collaboration among CSOs.

\section{Related Work}
This section situates our work at the intersection of civil society studies and research on online monitoring by positioning CSOs as distinct monitoring actors, reviewing the structural conditions shaping their work, and summarizing what is known and unknown about monitoring practices and workflows.

\subsection{CSOs as Distinct Actors in Online Monitoring}
\label{ssec:csos_distinct_actors}
Civil society organizations (CSOs) are non-state, non-profit actors pursuing normative political, social, or human rights goals, often under politically sensitive conditions \cite{osce_2018,capacity4dev_thematic_nodate}. 
When observing online discourse, they operate alongside actors from academia, journalism, fact-checking (FC), and content moderation (CM), but their roles, practices, and constraints differ substantially.

CSOs engage in long-term monitoring, evidence gathering, contextual analysis, reporting, as well as advisory and educational work. Further, their work can address local or rural contexts where far-right dynamics tend to be stronger \cite{diwBerlinGerman} but are often underreported by national media or neglected by academia. 
CSOs focus on sustained, contextual observation of sociopolitical dynamics; their profound knowledge of local dynamics, trends, and actors enables them to function as early warning mechanisms and provide timely, context-specific responses, while also connecting to regional and international CSO networks \cite{osce_2018,ihler_monitoring_2025}. 
This role becomes particularly salient as platforms and public institutions scale back regulation under the banner of free speech \cite{benavidez_big_2023, kayyali_metas_2025}. In this context, CSOs increasingly serve as both a supplement to and a corrective of institutional responses \cite{smidt_civil_2025}, challenging official narratives that may trivialize right-wing activity, ignore political motivations of violent acts, or deny broader authoritarian or racist undercurrents in society \cite{Bernhard2019,ihler_monitoring_2025}.

\subsection{Challenges Shaping the Working Conditions of CSOs}
\label{ssec:challenges}

Existing literature suggests that CSO monitoring practices are shaped by a combination of structural, political, and technological challenges. However, much of this knowledge stems from grey literature, including NGO self-reports \cite{hulshof-schmidt_2024_2024,noauthor_state_2025}, policy documents \cite{FRA2018,noauthor_protecting_2023}, and practitioner-oriented publications \cite{ihler_monitoring_2025}, rather than from systematic academic research. 

Across national contexts, CSOs face chronic financial uncertainty, largely due to reliance on short-term project funding \cite{hulshof-schmidt_2024_2024,noauthor_striving_2023,hummel_civil_2022}. This limits their capacity to invest in digital infrastructure, monitoring-specific tools, and dedicated technical staff \cite{hulshof-schmidt_2024_2024,hummel_civil_2022,noauthor_state_2025}. 
However, the consequences of these funding structures are not evenly distributed across CSOs. Larger, better-funded or digitally native organizations are better positioned to develop in-house monitoring tools \cite{isd_digital_2025,cemas_about_2025}, which are typically tailored to organizational needs and resources and not openly shared with others. This reflects the precarious situation on the `CSO-market', where organizations compete for highly limited funding \cite{hummel_civil_2022,noauthor_striving_2023,priemer_wie_2015}, creating incentives to emphasize differentiation over sustained collaboration, even in the presence of overlapping goals and values. 
At the same time, CSOs often operate in politically precarious environments, exposing them to harassment, intimidation, legal uncertainty, and targeted attacks by extremist groups, radicalized individuals, or state actors \cite{FRA2018,noauthor_protecting_2023,smidt_civil_2025}. 

Survey-based work highlights vulnerabilities related to security and privacy \cite{samarin_surveying_2020}, but also illustrates the difficulty of capturing CSO practices through standardized instruments. Differences in political focus, organizational structure, terminology, and regional context suggest that these challenges are highly situated and best addressed through qualitative, interactive, in-depth approaches, as we apply in our work. CSCW research on invisible work conceptualizes essential labor that remains hidden and informal \cite{star_layers_1999}. This lens helps situate civil society monitoring as a form of collaborative work outside institutionalized settings and often deliberately invisible due to legal and security concerns.

\subsection{Monitoring Practices and Workflows in CSOs and Adjacent Domains}
\label{ssec:practices_and_workflows}

Despite the important and distinct role of CSOs in online monitoring, direct empirical research on how they conduct their work remains scarce, with Bäumler et al.'s study \cite{baumler_harnessing_2025} providing the most related work to ours, offering important insights into the challenges of CSOs focused on processing user-submitted incident reports. 
To situate CSO monitoring practices, we draw on additional research from adjacent domains which provides a conceptual understanding for studying their monitoring work. 
This literature conceptualizes monitoring efforts as a ``pipeline of practices'' \cite{micallef_true_2022} involving proactive searching, filtering, interpretation, documentation, and prioritization of content \cite{zachlod_social_2021,evangelista_systematic_2021}. 
Studies of fact-checkers and journalists show that these pipelines are typically characterized by fragmented workflows, reliance on ad-hoc tools and substantial manual and repetitive effort \cite{micallef_true_2022, juneja_human_2022, beers_examining_2020,nordis_report_2022, nordis_policy_2022} 
Moreover, the access to tools is often limited while usability is described as rather low \cite{baumler_harnessing_2025}. 
The widespread lack of dedicated, monitoring-specific tools often forces practitioners to rely on makeshift solutions like spreadsheets, generic platforms \cite{beers_examining_2020} or Open Source Intelligence (OSINT) tools, initially developed for military and intelligence purposes, and therefore not necessarily aligned with the ethical values of CSOs \cite{nordis_report_2022}.

This proactive and investigative type of work is particularly labor-intensive, as emphasized in OSINT \cite{decary-hetu_sifting_2015, glassman_intelligence_2012,evangelista_systematic_2021} and FC research \cite{juneja_human_2022,micallef_true_2022}. 
At the same time, this stage remains relatively under-examined in research \cite{hrckova_autonomation_2024}, in contrast to later stages in workflows such as claim verification \cite{hrckova_autonomation_2024,nakov_automated_2021} or dissemination of insights \cite{hrckova_autonomation_2024,skopik_osint_nlp_2024}. 
This gap is consequential because it suggests that existing tools and system designs are often built on an incomplete account of this activity.
Further, current research provides limited insight into how monitoring workflows are enacted within CSOs' everyday  practices, under conditions of fragmented tooling and sustained manual effort, or how these conditions shape their expectations toward technological support. Within CSCW, such tool ecologies are understood as relational sociotechnical infrastructures sustained through ongoing maintenance rather than stable systems \cite{star_steps_1994}. Relatedly, CSCW research on bricolage highlights how actors improvise with heterogeneous resources to sustain work in constrained environments \cite{buscher_landscapes_2001,karanasios_digital_2022}.

\subsection{Technological Support in Monitoring Practices}
\label{ssec:technological_support}

An early and fundamental step in monitoring work is data collection, which is increasingly constrained by limited and fragile access to data due to the termination of affordable access to major platforms, the discontinuation of social media analytics tools \cite{ortutay_meta_2024}, and the restricted and opaque access to messaging platforms -- despite their growing relevance for the dissemination of harmful content \cite{barbosa_not_2019,nordis_policy_2022,hrckova_autonomation_2024}. These developments underscore the structural vulnerability of platform-dependent data access \cite{nordis_policy_2022}. 
Although legal exemptions have enabled the professionalization of academic data infrastructures \cite{gesis_telegram_toolkit}, CSOs rarely benefit because they lack comparable research privileges. Monitoring systems developed by established German CSOs \cite{cemas_about_2025, isd_digital_2025} are also not (per se) accessible to others. Meanwhile, open-access data collection and analysis tools \cite{primig_introducing_2024,silva_telegramscrap_2024,ruscica_telecatch_2025,peeters_4cat_2022} are often technically inaccessible, poorly maintained \cite{schinas_open-source_2017}, and rarely informed by systematic engagement with practitioners. CSCW research on infrastructuring emphasizes that systems evolve through and against an ``installed base '' \cite{neumann_making_1996} and are continuously maintained rather than ever completed \cite{karasti_infrastructuring_2014}.

Research has confirmed the importance of supporting what has been described as ``low-tech needs'' in investigative and monitoring work \cite{hancock_seeing_2024}, including steps such as reviewing search results, identifying relevant material, and extracting key information \cite{mckeown_investigating_2014}.
At the same time, recent research has increasingly focused on developing `cutting-edge' machine learning-based systems, such as tools for detecting, classifying, and organizing harmful online material in CM \cite{seering_chillbot_2024,schluger_proactive_2022} and FC contexts \cite{liu_human-centered_2024,hrckova_autonomation_2024,das_state_2023,guo_2024}. In principle, such systems appear potentially relevant for CSO-based monitoring, as they promise support for identifying large volumes of potentially relevant material. In investigative and OSINT settings, tools combining NLP techniques, such as named entity recognition, search, filtering, and AI-assisted reporting, have been shown to support early-stage analysis and content triage \cite{skopik_osint_nlp_2024}. 
Monitoring workflows may further require tools to process and analyze multimodal data, such as images or audio, as suggested by research in adjacent domains \cite{cifliku_save_2025,wolfe_implications_2024}.

However, the practical utility of current learning-based systems for CSO monitoring remains uncertain. Many moderation tools and training datasets are proprietary and inaccessible, and it is often unclear whether existing tools align with the concepts, categories, and contextual sensitivities of CSOs. Locally embedded and evolving phenomena -- such as neo-fascist youth rhetoric in rural Germany -- are unlikely to be captured by general-purpose hate speech classifiers built around platform policies or legal thresholds. From a CSCW perspective, this highlights the difficulty of translating locally embedded monitoring practices into standardized computational categories, echoing CSCW work on (heterogenous) knowledge production and boundary objects \cite{star_institutional_1989,star_steps_1994,ackerman_sharing_2013}.

Technical limitations persist, including degraded performance outside high-resource languages and standard settings \cite{koenecke_etal_2020,nordis_policy_2022}. Even widely used tools under-detect implicit hate speech, over-moderate counterspeech \cite{Hartmann_2025}, and fail to capture language- and context-specific patterns \cite{nozza_exposing_2021,kovacs_challenges_2021}.
Moreover, even effective systems require substantial human correction and careful workflow integration, including appropriate interfaces \cite{skopik_osint_nlp_2024}.

Despite these limitations, practitioners in FC and CM have generally expressed openness toward AI-based tools primarily for narrow, well-defined tasks rather than for complex interpretive or evaluative work \cite{liu_human-centered_2024,nordis_policy_2022,hrckova_autonomation_2024}. 
In these interventionist contexts, where tools are designed to support enforcement, removal, or feedback through detection and flagging  \cite{schluger_proactive_2022,seering_chillbot_2024,bozarth_wisdom_2023}, human-AI collaboration is often framed as a way to address the limitations of automated detection, for example via conditional delegation \cite{lai_human-ai_2022}.
Monitoring, by contrast, can be understood as a form of epistemic work that prioritizes observation and sensemaking over intervention, suggesting different expectations toward automation and a stronger emphasis on assistive rather than decision-making tools.

Prior research highlights a preference for human-in-the-loop systems that balance algorithmic scale with human control, especially in high-stakes or interpretive contexts \cite{demartini_human_loop_2020,liu_human-centered_2024,nakov_automated_2021}. Practitioners often exhibit ``algorithm aversion'' \cite{liu_human-centered_2024}, driven by concerns about trust, transparency, and reliability \cite{nguyen_believe_2018,hrckova_autonomation_2024}. As a result, recent work emphasizes participatory design \cite{hancock_seeing_2024} and flexible, modular tools that support transparency, adjustability, and user control \cite{nakov_automated_2021,nordis_report_2022,das_state_2023,wolfe_implications_2024,baumler_harnessing_2025,schopke_ots_2025}, framing AI as collaborative support rather than automation \cite{morris_is_2024}.

Existing CSCW research suggests positioning CSO monitoring as a form of collaborative, infrastructural, and epistemic work. Despite this, existing accounts of monitoring practices largely derive from journalism, FC, or intelligence settings and offer limited insight into how such work is enacted in long-term, locally embedded, and normatively driven civil society contexts. While fragmented tools and high manual effort are well documented, less is known about how CSOs navigate technical, ethical, and legal constraints or how these shape expectations for technological support. This study addresses these gaps by examining CSOs' monitoring practices (RQ1), challenges (RQ2), and needs for technological support (RQ3).

\section{Methodology}
\label{sec:methods}
This study was conducted as part of a research-to-practice project in collaboration with a partner CSO.\footnote{https://arai-telegram.github.io/} In parallel to the empirical work, an open-source prototype was developed and iteratively adapted over the course of the project. The project sought to integrate CSOs' requirements into the design and development process, and this study provides the groundwork for defining and prioritizing key software features.

While the technical focus of the software is on Telegram, a key platform for far-right actors in Germany and beyond \cite{baumgartner_pushshift_2020,urman_what_2022}, the aim of the study is broader: to gain a realistic understanding of the workflows, challenges, and requirements of CSOs engaged in online monitoring. This includes gathering insights that are transferable to other social media platforms, services, and communication modalities. In the interviews, we therefore asked participants to reflect on other platforms and services as well.

\subsection{Participants}

To gain in-depth and contextual insights, we conducted qualitative interviews without offering financial incentives (see Section \ref{sec:research_ethics}). Access to participants was facilitated through existing contacts in civil society, including connections established via the project's partner CSO. Participants were selected through non-probability sampling, combining purposive (expert-based) \cite{palinkas_purposeful_2015} and snowball methods \cite{etikan_2015,galloway_2005}. We spoke with individuals and representatives from 12 CSOs engaged in monitoring, education, archiving, analysis and publication of insights, as well as communication- and security-related consulting including threat assessment for individuals and other civil society organizations. Their activities can be broadly categorized into three primary foci:
\textit{Archiving} -- systematically collecting material to preserve insights into far-right activities, creating resources for researchers and journalists.
\textit{Monitoring} -- documenting incidents, tracking narratives, symbols, and emerging events, and providing both qualitative and quantitative analysis of trends and threats.
\textit{Counseling/Advisory} -- offering threat assessments and guidance to individuals or groups targeted by far-right actors, helping communities respond to potential risks.
Three participants had educational backgrounds in technology, which in some cases influenced how they articulated ideas about potential tool functionalities, demonstrating greater awareness of technological possibilities. 
Table \ref{tab:participants} provides an overview of the participants’ organizational focus, main activities, and technical backgrounds. Participant-level information is available in the Appendix (Table \ref{tab:participants_detail}).

\begin{table}[h!]
\caption{Overview of interviewed CSOs and participant roles (n=15).}
\label{tab:participants}
\centering
\small
\begin{tabular}{lp{0.75\textwidth}}
\toprule
\textbf{Dimension} & \textbf{Distribution} \\
\midrule
Organizational focus &
Far-right (8), Antisemitism (2), Conspiracy theories (3),  Racism (1), 
Human rights violations (1), Various phenomena (3) \\

Main activity &
Monitoring (7), Consulting (5), Archiving (2), Analysis (2), Education (1), Software development (1) \\

Technical education &
Yes (3), No (12) \\
\bottomrule
\end{tabular}
\end{table}

\subsection{Interviews}
Between October 2024 and January 2025, we conducted 13 interviews with a total of 15 participants. These took place remotely via a video-conferencing tool, allowing for more flexible scheduling and participation from different regions. Most interviews were conducted with one person; in two cases, two participants from the same organization took part together. Interviews lasted between  30 minutes and one hour and were led by one of the authors using a semi-structured guideline. A translated version of the interview guide is included in the Appendix (\ref{ssec:interview_guidelines}). 

Each interview began with an introductory question inviting participants to describe themselves and their organization's work. The conversation moved through thematic areas such as current practices and challenges in their monitoring work, relevant platforms and content types, tools used for monitoring, perceptions of current technologies, including AI, and their potential for enhancing monitoring work. Follow-up questions were adapted to each conversation. Since participants had varied levels of technical knowledge, interviewers sometimes introduced specific examples -- such as topic modeling, clustering, network analysis, or video summarization -- to help illustrate possible technological support and spark discussion. 

Interviews were audio-recorded when possible. In two cases where participants preferred not to be recorded, a second researcher attended and produced a detailed written protocol, occasionally asking clarifying follow-up questions.

\subsection{Research Ethics and Anonymization}
\label{sec:research_ethics}
Studying CSOs that conduct monitoring of antidemocratic content and actors poses a distinctive ethical challenge with the risk of increasing the exposure of these organizations to the very actors they monitor, and careful consideration must be taken to avoid revealing identifiable details about their identity, methods, and vulnerabilities. We therefore drew on guidelines for Internet Research Ethics \cite{franzke_internet_2020} and research on online terrorism, extremism, and far-right actors \cite{conway_online_2021,massanari_rethinking_2018}. Prior to the interviews, the participants received detailed information about the study's objectives, data collection procedures and data handling practices. All participants provided informed consent to participate in the study. Consent to audio recording was obtained separately and was not a requirement for participation. Participants were assured of strict confidentiality, and all shared information was anonymized to protect both individual and organizational identities. 
Moreover, because relatively few organizations engage in this work, even indirect identifiers, such as the region in which an organization operates, the specific groups they monitor, or their founding date, could inadvertently reveal their identity. Therefore, we refrained from describing participants or their CSOs in detail.

Interviews were automatically transcribed using Whisper\citet{open_ai_introducing_2022} operated locally. Following manual proofreading of the transcripts, all audio and video files were securely deleted in accordance with the consent information provided to participants. Any identifiable details, such as the names of participants and organizations, were removed from the transcripts. Anonymized transcripts were encrypted and stored with password protection to ensure secure data management.

Participants were not financially compensated for their participation. Most took part as part of their regular job responsibilities during working hours. Additionally, participants expressed interest in shaping the software tool being developed through the research project, aligning their involvement with their professional goals. The online format of the interviews also eliminated any potential out-of-pocket expenses, such as travel costs.

\subsection{Researcher Positionality}
The mediated access to interview partners through the partner CSO fostered trust and supported an open interview atmosphere, particularly given the sensitivity of the topic and the precarious position of many CSOs. At the same time, participants were aware of the researchers' involvement in the development of AI-based tools for the automated detection of hateful content. This awareness may have shaped how certain issues were framed, emphasized, or problematized during the interviews. Importantly, however, participants frequently articulated critical and skeptical views toward automated detection systems, suggesting that responses were not primarily oriented toward affirming the interviewer's assumed research agenda, but realistically reflected participants' situated practices and concerns.

\subsection{Data Analysis}
We employed Thematic Analysis \cite{braun_using_2006}, an established approach in CSCW (e.\,g. \cite{seering_chillbot_2024,schluger_proactive_2022,mukhopadhyay_osint_2024}), which allows for the identification of key patterns and themes within qualitative data, providing a rich understanding of participants' perspectives and experiences. While the research questions structured the interview guidelines and thus also the data collection, the data analysis itself was inductive, data-driven. 

Familiarization with the data, the first step of the thematic analysis, was conducted by the first author. While reviewing generated transcripts against the audio recordings, notes from interviews that were not audio-recorded were integrated and initial analytic notes added. Transcripts were organized in a table format to support annotation at the level of individual statements.

For the initial coding phase we employed an open coding approach. Meaningful chunks of data -- ranging from single sentences to short passages -- were assigned codes that stayed close to participants' wording and focused primarily on content. To establish a shared understanding of the data and the coding approach, the first and second author independently coded two interviews. The resulting codes showed strong correspondence and were discussed to align interpretations. Subsequently, the first author coded all remaining interviews and developed a preliminary code tree. 

In a process resembling axial coding, the first author grouped low-level codes into higher-level categories and themes, moving toward more interpretive abstractions. For example, statements describing the time-consuming and non-scalable nature of manual monitoring practices were initially coded descriptively (e.\,g., as labor-intensive monitoring) and later grouped into broader themes capturing the initial situation of monitoring work. 
This process resulted in a preliminary thematic map comprising 714 codes organized in 25 themes.
The process of searching for and reviewing themes was iterative and collaborative. Both authors reviewed and discussed the themes and subthemes, reconsidering their naming and reassigning certain codes where necessary. These discussions informed the revision of the first thematic map, yielding 21 themes and 497 codes.

Throughout this process, a shared digital platform was used to collaboratively organize and visualize codes and themes, supporting the identification of overlaps and conceptual relationships. The process was continued until the final themes were ``deemed internally coherent, consistent and distinctive'' \cite{braun_using_2006}, reflecting the analytical depth and clarity of the thematic map.

Finally, the stabilized themes were mapped onto the research questions to support structured reporting of the findings. Some themes spanned multiple clusters and were relevant to more than one research question. For example, the manual effort involved in monitoring practices relates both to current workflows and to demands for enhanced technological support, and therefore informs multiple research questions. For reporting purposes, such themes were assigned to the section where they were most salient, based on the authors' judgement.  Importantly, this mapping occurred after the thematic structure had been established and should be understood as an analytic organization of results rather than a predefined coding scheme. 

The data analysis ultimately consolidated into four meta-themes, 21 themes, 80 subthemes, and 329 codes that were finally translated into English. Four meta-themes form the structural backbone of the Findings section: Current Practices and Workflows (\ref{ssec:current_practices}), Technical, Ethical and Legal Challenges (\ref{ssec:rq_challenges}), Needs and Expectations Regarding Technological Support (\ref{ssec:needs_expect}), and Perspectives on the Role of AI and Human-AI Collaboration (\ref{ssec:human_ai}).
For the presentation of the findings, we selected vivid and representative examples from the transcripts to illustrate the identified themes and subthemes. 

\subsection{Workshops}
In addition to the interviews, we conducted two one-day participatory workshops with interview participants as a formative co-design and exploratory evaluation step. The goals were to ground the conceptual monitoring workflow in practitioners' everyday monitoring practices and to gather situated feedback on an early prototype of the Telegram monitoring tool.
Prior to each workshop, participants were invited to submit Telegram channels relevant to their ongoing work. These were pre-integrated into the system to support realistic hands-on exploration. Each workshop combined a presentation of key findings from the interviews and the emerging monitoring workflow, hands-on tool use centered on participants' own thematic interests, and moderated group discussions.
During the hands-on sessions, participants registered user accounts, tested lexical and semantic search, inspected automatically generated audio transcripts, and explored a ``conspiracy score'' on message level provided by an automated classifier.  
We documented the session through notes, which informed the iterative refinement of the workflow and the prototype. 

\section{Findings}
\label{sec:findings}
Participants report working under constrained financial, personnel, and technical conditions, consistent with prior research on CSOs \cite{samarin_surveying_2020,baumler_harnessing_2025}. Limited technical expertise constrains their use of digital tools. At the same time, they stress that online spaces are central to their work, as antidemocratic actors increasingly exploit digital platforms and online harms translate into real-world consequences, reinforcing existing research \cite{eu_parliament_2024,adl_hate_report_2023,conway_right-wing_2019}. Across interviews, this underscores the urgency of effective online monitoring, improved workflows, and more accessible tools.
In the following, we address our research questions in detail and situate our findings within existing research. We include figures to show the relevant themes and subthemes per section. While codes are largely collapsed for readability, node labels indicate the number of codes. A full code list is available via Zenodo.\footnote{https://zenodo.org/records/20447764}

\subsection{Current Practices and Workflows (RQ 1)}
\begin{figure}
    \centering
    \includegraphics[width=\linewidth]{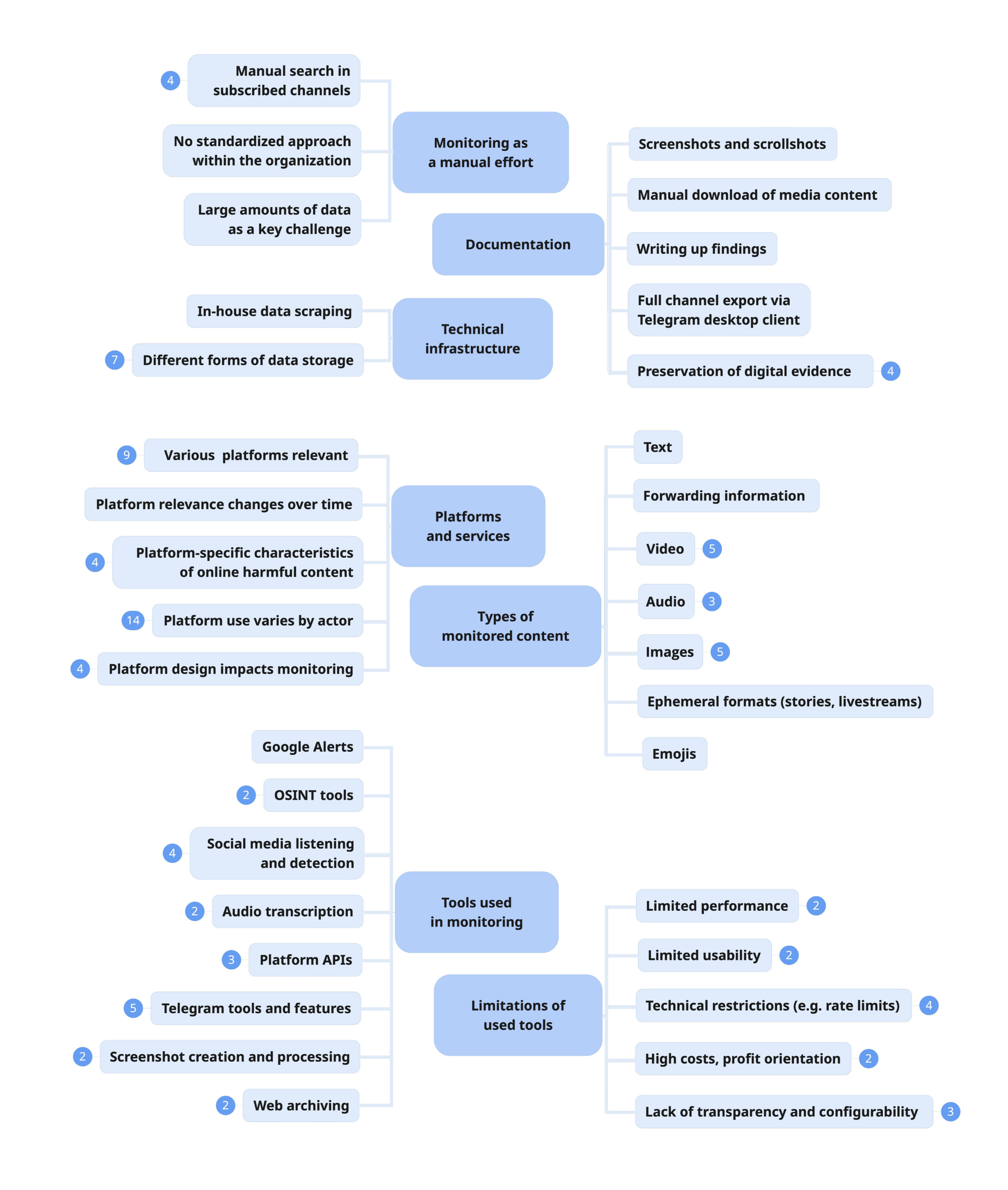}
    \caption{Thematic map section of CSO monitoring practices and workflows. Themes (blue nodes), subthemes (light blue nodes), and code counts (collapsed for space) are shown. White labels indicate individual codes not assigned to a subtheme.}
    \label{fig:current}
    \Description{Thematic map section of CSO monitoring practices and workflows. Themes (blue nodes), subthemes (light blue nodes), and code counts (collapsed for space) are shown. White labels indicate individual codes not assigned to a subtheme. The map was created based on qualitative coding of 15 interviews with CSO practitioners working for German-speaking organizations.}
\end{figure}

Figure \ref{fig:current} illustrates the themes and subthemes characterizing current CSO monitoring practices and workflows.

\label{ssec:current_practices}

\subsubsection{Monitoring as a Manual Effort}
Consistent with \citet{baumler_harnessing_2025}, participants describe online monitoring as largely manual, carried out routinely or in response to offline events. They rely e.\,g. on the Telegram desktop client or mobile app to follow channels and search content, and have to``trigger the full-text search manually every time'' (P4). Automation via the Telegram API is rarely used, and documentation is likewise done by hand -- typically through screenshots and notes -- making the process ``very time-consuming and ineffective'' (P11). This mirrors findings from research on investigative journalism and FC workflows \cite{beers_examining_2020,das_state_2023,juneja_human_2022}.

\subsubsection{Documentation of Monitoring Insights}
Participants rely on largely low-tech methods to document insights, most commonly screenshots, manual media downloads, and written notes. As P6 explains, documentation often ``starts with screenshots -- and then, if necessary, we write it down separately.'' For longer discussion threads, scrollshots are used, though participants note the lack of convenient desktop tools, especially for Telegram comment threads, where ``you end up with a hundred photos'' (P6). Some participants also use the Telegram desktop client's export function to archive entire channels, which is seen as ``very helpful,'' particularly when filtering by date or media type (P6).
Screenshots are described as essential evidence in cases involving criminal content, where ``the incitement to violence and the account must be identifiable'' (P9). This aligns with prior work identifying screenshots as a standard OSINT documentation practice \cite{mckeown_investigating_2014,baumler_harnessing_2025}. Notably, only one participant raises concerns about evidence manipulation. To ensure authenticity, they employ cryptographic hashing and independent web archiving e.\,g. via the Internet Archive \cite{internet_archive_2025} to establish a ``chain of evidence'' and ``prove that things weren’t tinkered with afterwards'' (P13), reflecting a level of technical rigor rarely noted in prior research.

\subsubsection{Technical Infrastructure}
Participants report varied data storage practices, most commonly relying on local file storage for screenshots, links, and text files. These practices are often ad hoc and unsystematic, particularly for newer content formats. Few organizations maintain databases, which are typically manual and text-only, and only two participants report using in-house scraping scripts. Overall, this points to a lack of technical infrastructure for systematic data management, echoing findings from journalism research \cite{beers_examining_2020} and reflecting what CSCW describes as relational infrastructures sustained through ongoing local maintenance rather than stable systems \cite{star_steps_1994} and evolving  through an installed base \cite{neumann_making_1996,star_steps_1994}.

\subsubsection{Relevant Platforms and Services}
Participants identify mainstream platforms (Instagram, TikTok, Facebook, YouTube, X) and Telegram as most relevant, with some also monitoring VKontakte, right-wing forums, imageboards, and comment sections of media outlets. Platform relevance shifts over time: as P7 notes, ``Facebook [was relevant] in the past'' , while Telegram ``is still relevant to a certain extent, but there is not so much going on publicly anymore''. Newer platforms can disrupt established workflows -- TikTok, for instance, lacks channel structures that can be easily archived (P8).
Participants also describe platform-specific differences in the prominence of various phenomena: Instagram and TikTok as particularly important for right-wing extremism among younger audiences (P11), while ``stalking takes place on Instagram, and hate campaigns take place on X'' (P4). In line with prior research \cite{rogers_propagation_2023}, participants emphasize the importance of cross-platform analysis to obtain a ``seriously complex picture'' (P4).

\subsubsection{Formats and Modalities of Monitored Content}
Participants stress that effective monitoring must address the increasingly multimodal nature of harmful content, in line with FC and journalism research \cite{nordis_report_2022,cifliku_save_2025}. 
Text remains the primary focus, as it yields substantial insight while being comparatively efficient and reliable to analyze (P14). 
Visual content is considered more resource-intensive to store and process, but also important: Participants mention sharepics, videos from demonstrations, and emojis used as coded signals. 
The relevance of audio formats varies: while voice messages have declined in relevance, podcasts and long-form livestreams -- often recorded on video -- remain valuable, particularly for identifying actors and documenting statements made at events (P9).

\subsubsection{Tools Used in Monitoring}
While monitoring is largely manual, some participants use tools to support specific tasks, often repurposing services from commercial social media listening or OSINT. For search and discovery, web search and Google Alerts are commonly mentioned. Web search and OSINT tools reflect practices observed in FC \cite{nordis_report_2022}. Use of social media platform APIs is rare, as access is limited or no longer available \cite{ortutay_meta_2024}.
To analyze audio content, participants rely on existing podcast or YouTube transcripts or experiment with the open-source ASR model Whisper \cite{open_ai_introducing_2022}. For image content, free and easy-to-use tools such as PDF readers are used to search screenshot-based PDFs. For documentation, some CSOs rely on web archives, a practice also noted in FC \cite{mckeown_investigating_2014,nordis_policy_2022}.
Telegram-specific (paid) features include built-in translation, channel recommendations, analytics services (e.\,g., telemetr.io), and custom bots for keyword alerts. A small number of participants use or plan to use a paid Telegram monitoring service or automated hate speech detection tools. Monitoring is primarily conducted in German, with translation needs handled via built-in tools and seen as helpful but not essential, unlike in reporting-focused CSOs \cite{baumler_harnessing_2025}. This may reflect that monitoring-focused CSOs actively select content, whereas reporting-focused CSOs process user submissions.
Taken together, these practices reflect a form of bricolage \cite{buscher_landscapes_2001,karanasios_digital_2022}, in which monitoring work is assembled through the opportunistic recombination of heterogeneous tools rather than through integrated systems, producing what CSCW describes as relational infrastructures sustained through local configuration and ongoing maintenance \cite{star_steps_1994}.

\subsubsection{Limitations of Tools in Use}
Participants report multiple tool-related limitations, including the cost of subscription services and technical issues such as API limits and errors. Workflows are often cumbersome, for example requiring manual audio downloads followed by self-scripted ASR processing.
Proprietary tools are criticized for limited transparency and configurability, reducing their value for research and investigative work. Participants also express concern about error rates in AI-based classification  -- particularly for threat assessment, where a high recall is critical. 
Finally, some highlight biases linked to commercial business models that prioritize easily prosecutable cases and specific platforms, especially Facebook. P4 criticizes that such services ``place their bet on the winning horse and, as a result, become kind of one-dimensional''. Platform biases have been observed in FC workflows as well, which are shaped by platform-provided infrastructure and funding \cite{nordis_policy_2022,belair-gagnon_knowledge_2023,beers_examining_2020}.

\subsection{Technical, Ethical, and Legal Challenges (RQ 2)}
\label{ssec:rq_challenges}

\begin{figure}
    \centering
    \includegraphics[width=\linewidth]{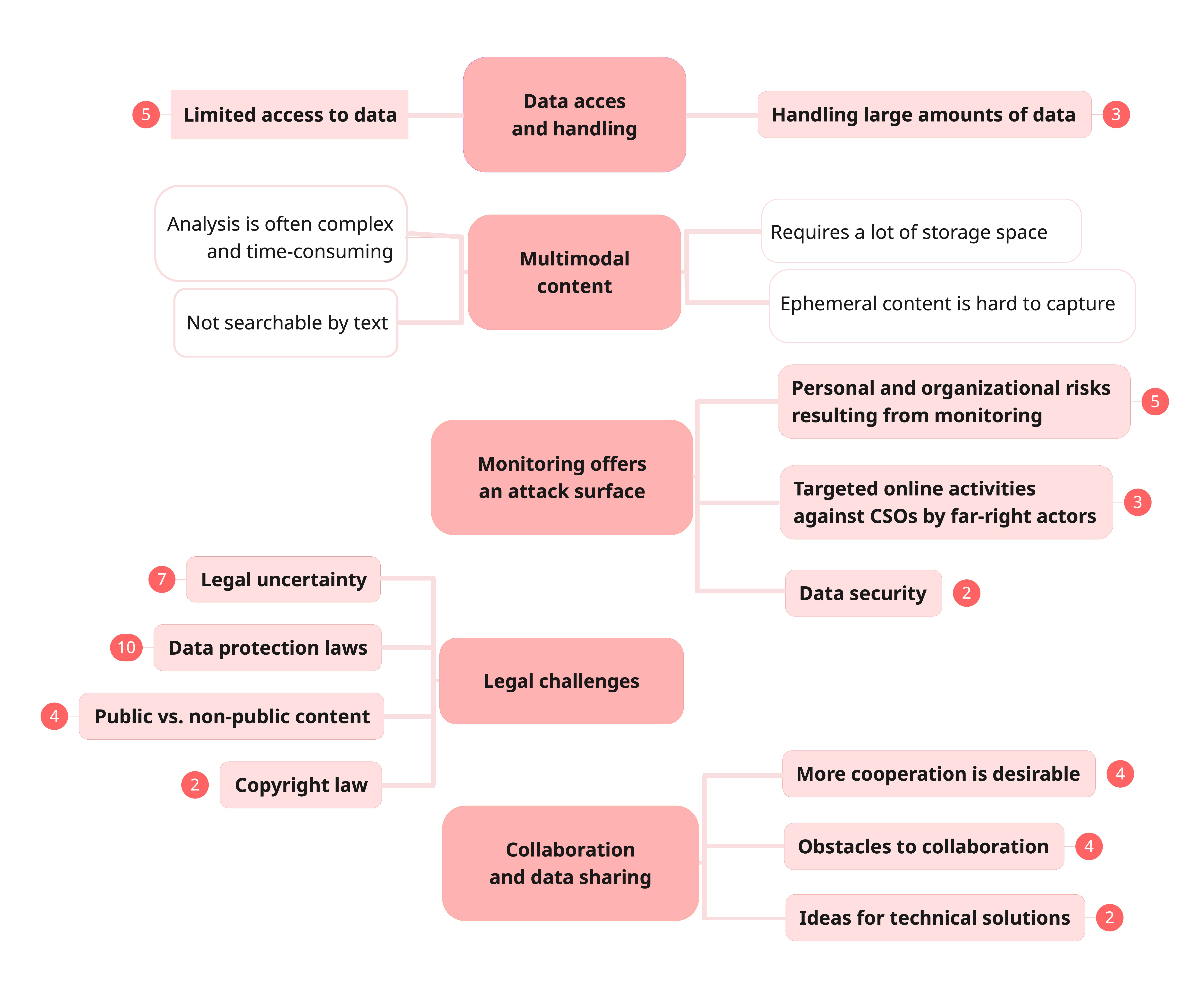}
    \caption{Thematic map section of technical, ethical and legal challenges of CSOs in monitoring work. Themes (red nodes), subthemes (light red nodes), and code counts (collapsed for space) are shown. White labels indicate individual codes not assigned to a subtheme.}
\label{fig:challenges}
 \Description{Thematic map section of technical, ethical and legal challenges of CSOs in monitoring work. Themes (red nodes), subthemes (light red nodes), and code counts (collapsed for space) are shown. White labels indicate individual codes not assigned to a subtheme. The map was created based on qualitative coding of 15 interviews with CSO practitioners working for German-speaking organizations.
 }

\end{figure}

Figure \ref{fig:challenges} summarizes the key challenges practitioners face in monitoring, along with the associated themes and subthemes.

\subsubsection{Data Access and Handling}
\label{sssec:data_access_handling}
Participants report major challenges related to limited data access due to platform policies. While Telegram data is described as relatively accessible, access to Instagram or Facebook is ``way more difficult'' (P6). Participants also highlight changes in API access at Meta and X, noting that previously accessible data has become unavailable or prohibitively expensive for research (P13, P14). Similar dependencies and disruptions have been widely documented across sectors, as key tools and APIs have been discontinued or monetized \cite{baumler_harnessing_2025,hancock_seeing_2024,hrckova_autonomation_2024,juneja_human_2022,nordis_report_2022,beers_examining_2020}.
Beyond access restrictions, participants struggle with data volume. Social media is described as a ``bottomless pit'' (P8), where large-scale collection becomes unmanageable without adequate analytical tools. Some participants report compute limits, with large datasets overwhelming local resources and causing crashes. Monitoring-focused CSOs thus face greater constraints than FC or CM workers, or reporting-oriented CSOs, who benefit from more established infrastructures or curated inputs \cite{juneja_human_2022,hrckova_autonomation_2024,belair-gagnon_knowledge_2023,baumler_harnessing_2025}.

\subsubsection{Multimodality of Content}
\label{sssec:challenge_multimodal}
Another major challenge stems from the multimodal nature of monitored content. Audio and video are described as especially time-consuming, as they cannot be keyword-searched and require full manual review: ``you have to look through it completely to find the relevant parts'' (P2). While such formats can yield important insights, participants lack the time and tools to analyze hours-long recordings.
Accordingly, almost all participants highlight the value of transcription tools for audio and video, echoing prior research \cite{cifliku_save_2025,juneja_human_2022,liu_human-centered_2024}. Beyond this, participants point to ephemeral formats such as livestreams and stories as a distinct challenge due to their limited availability. Some suggest that malicious actors deliberately use these formats, which are ``difficult to capture'' and thus less likely to be monitored or archived effectively (P14).

\subsubsection{Legal Challenges}
\label{sssec:legal_challenges}

Many participants describe their monitoring work as shaped by substantial legal uncertainty, particularly because they operate ``like journalists, but are not journalists'' (P15). This creates a dilemma: they must retain evidence to substantiate their public statements and defend themselves against potential legal challenges, while data protection and GDPR limit what they are permitted to store. Similar challenges have been noted for reporting CSOs lacking a formal monitoring mandate \cite{baumler_harnessing_2025}. From a CSCW perspective, this positions monitoring as a form of invisible work \cite{star_layers_1999}, where essential evidentiary and protective labor remains institutionally unsupported and largely hidden from public view.
Compliance with data protection law is often difficult and, in some cases, leads organizations to scale back or stop monitoring altogether. Others restrict themselves to public content only. However, this limits insight, as threats and plans for violence often emerge in non-public spaces, a tension also discussed in prior research \cite{conway_online_2021}.
Finally, copyright restrictions further constrain collaboration, as collected material often cannot be shared across organizations and may only be shown indirectly.

\subsubsection{Security Concerns}

Beyond legal challenges, participants report significant security risks at both individual and organizational levels. Monitoring far-right actors can make practitioners and their organizations targets of harassment or attack.
At the individual level, participants describe using disposable or mock accounts for monitoring due to personal safety concerns, allowing profiles to be abandoned if exposed, a practice also noted in OSINT research \cite{mckeown_investigating_2014}. Personal security concerns were also evident during interview preparation, as some participants declined recording or emphasized the need to remain unidentifiable. Similar risks have been documented for academic researchers studying far-right actors online \cite{massanari_rethinking_2018,conway_online_2021}, though few studies address these issues for CSOs, which are often particularly vulnerable due to limited resources and frequent targeting \cite{samarin_surveying_2020}.
Participants further stress that monitoring can endanger entire organizations. As P1 notes, they are ``also being attacked,'' (P9) often through coordinated campaigns. Some report strategic misuse of legal instruments, such as calls to ``flood projects like us with GDPR requests,'' (P15) while others point out that ``far-right actors are very eager to sue'' (P3). These tactics mirror reports from reporting-focused CSOs, who describe deliberate attempts to overload their capacity \cite{baumler_harnessing_2025}.
Finally, participants express concern that a shifting political climate may heighten the risk of legal persecution. In this context, secure data storage is seen as critical, as monitoring work could be reframed by authorities as illegal data collection or the creation of ``enemy lists'' (P9).

\subsubsection{Collaboration and Data Sharing} 
\label{sssec:collaboration}

Legal constraints also hinder inter-organizational collaboration. As P2 explains, organizations cannot share direct database access ``for data protection reasons'' and must rely on anonymized material -- ``enormous amount of work'' that must be done manually by trained staff. A lack of standardization and limited resources further impede collaboration; national attempts at uniform documentation have ``regularly failed'' (P15) due to the excessive effort required. From a CSCW perspective, these proposals point to infrastructuring challenges -- building shared arrangements and boundary objects that can be maintained and governed across organizations \cite{karasti_infrastructuring_2014,star_institutional_1989}.
Despite these barriers, participants strongly emphasize the value of collaboration. Successful partnerships -- such as with reporting-focused CSOs -- are described as effective due to a clear division of labor. Others note that collaboration could reduce redundant work, though implementation remains challenging. Suggested solutions include shared infrastructures, such as a collaborative database, or partnerships where one organization ``captures the online context'' (P2) while others contribute complementary expertise. This hints at collaboration as a key asset beyond technical tools \cite{juneja_human_2022}, while echoing prior work that also documents its challenges \cite{baumler_harnessing_2025}.

\subsection{Needs and Expectations Regarding Technological Support  (RQ 3)}
\label{ssec:needs_expect}

\begin{figure}
    \centering
    \includegraphics[width=\linewidth]{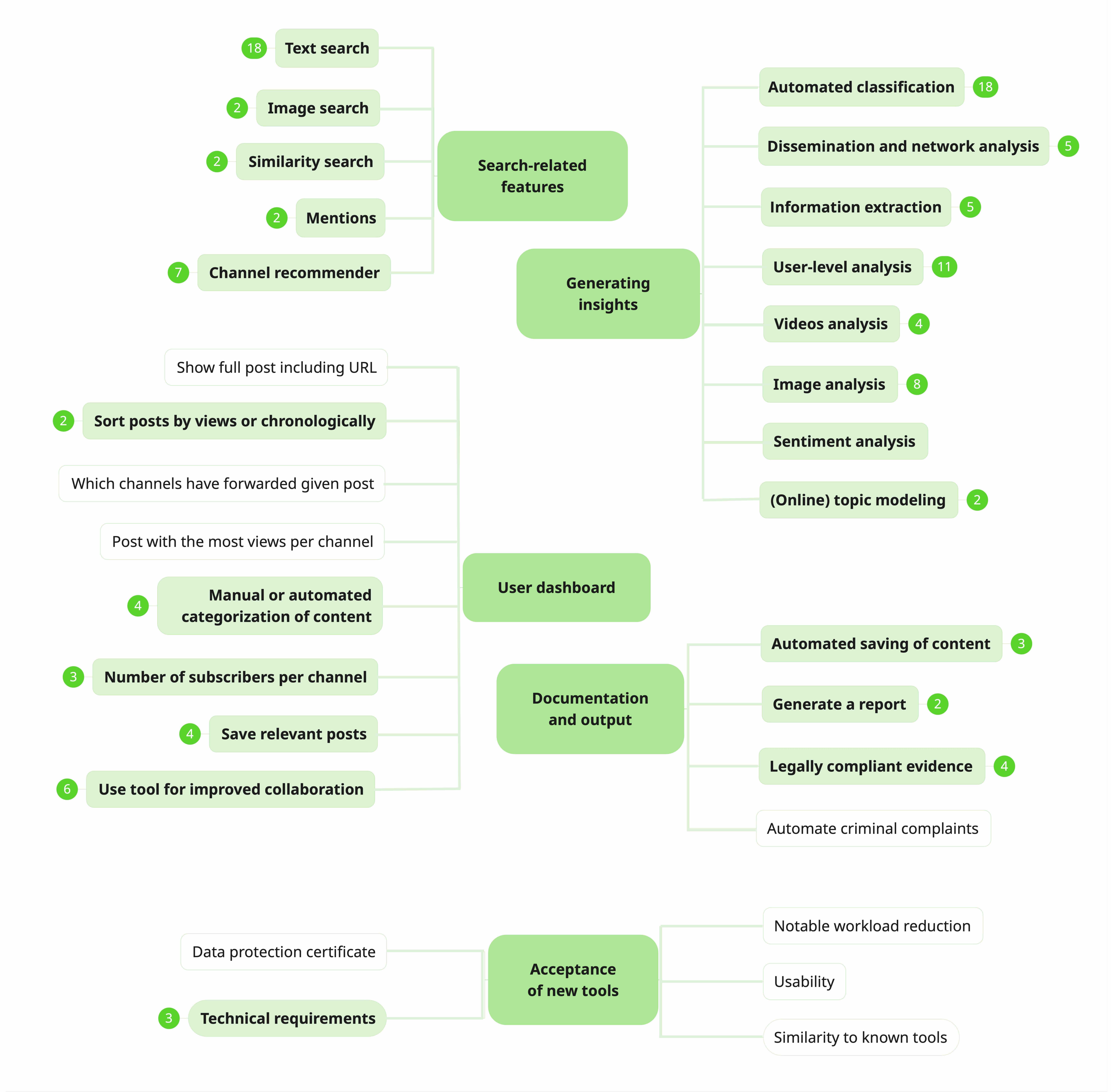}
    \caption{Thematic map section of CSO needs and expectations regarding technological monitoring support. Themes (green nodes), subthemes (light green nodes), and code counts (collapsed for space) are shown. White labels indicate individual codes not assigned to a subtheme.}
    \label{fig:demands}
 \Description{Thematic map section of CSO needs and expectations regarding technological monitoring support. Themes (green nodes), subthemes (light green nodes), and code counts (collapsed for space) are shown. White labels indicate individual codes not assigned to a subtheme. The map was created based on qualitative coding of 15 interviews with CSO practitioners working for German-speaking organizations. 
 }
\end{figure}

Figure \ref{fig:demands} presents participants’ technical needs for monitoring, clustered into four areas: search, interfaces and dashboards, analytical insights, and documentation. A fifth category addresses adoption factors such as usability, data protection, and local deployment.

\subsubsection{Enhanced Search and Discovery} 
\label{sssec:enhanced_search}Most participants identified advanced search capabilities as a central requirement for effective monitoring. A key demand was the ability to search beyond already known or subscribed channels. As P4 explains, they need ``a global search [...] not just looking in the groups we already know,'' ideally including ``full-text search in the entire Telegram''. Participants envision alert-based systems similar to Google Alerts, enabling highly targeted queries using advanced search techniques or ``Dorks'' (P4).
This emphasis on search and discovery aligns with prior research on OSINT analysts, investigative journalists, and fact-checkers, for whom search constitutes a core activity \cite{mukhopadhyay_osint_2024,mckeown_investigating_2014,beers_examining_2020,juneja_human_2022}.
Participants further stress the need to make multimodal content searchable, particularly audio and video. Access to transcripts is seen as critical, as ``a lot happens there that we don't really have on our radar yet'' (P14). From a technical perspective, this highlights the relevance of integrating automated speech recognition (ASR) and optical character recognition (OCR) as pre-processing steps.
Beyond expanding searchable content, participants emphasize automation, including saved queries and continuous monitoring. P4 describes the need for ``an automated search that runs [...] 24/7,'' while alerts are viewed as essential for surfacing relevant findings in real time, echoing similar demands in journalism \cite{cifliku_save_2025}. Search is also used to track mentions of known individuals or organizations that may become targets of attacks, for example through automated keyword searches across channels.
Finally, participants extend the notion of search to exploratory discovery, noting that ``sometimes you don't even know what you’re looking for'' (P14). Suggested features include similarity-based search as well as tools for identifying new and locally relevant channels as ecosystems shift over time. While Telegram's existing `Similar Channels' feature was seen as limited and repetitive, participants expressed interest in more transparent and context-sensitive channel recommendation mechanisms. 
These ideas provide novel insights and extend existing knowledge on search strategies in online monitoring by moving beyond keyword-driven approaches and highlighting the importance of channel discovery alongside content discovery.

\subsubsection{User Dashboard}  
\label{sssec:dashboard}
Participants propose a dashboard-centric interface for viewing and interacting with monitored content. Core requirements include a feed-based presentation of results with sorting and filtering option, alongside simple, accessible statistics, e.\,g. basic post-level metrics to identify the most-viewed posts within a channel.
Suggested metrics at channel level include channel growth and forwarding patterns, which, as P14 notes, can signal newly emerging channels and support exploratory monitoring. These dynamics are also linked to offline events, as demonstrations often coincide with spikes in follower numbers, helping to ``keep an eye on followers in the dynamics'' (P14).
Participants further envision the dashboard as a workspace for documentation and collaboration. P4 describes quickly saving relevant posts ``by clicking on a post and then it is immediately stored'' for later review or reporting. Dedicated user accounts could support standardized workflows and collaborative work, enabling users to scroll, flag items, and generate reports.

\subsubsection{Generating Insights}
\label{sssec:generate_insights}
Alongside search and dashboard-based work, the generation of insights occupies a substantial part of the thematic map. A central theme is automated content classification, which participants view as useful for pre-filtering and prioritizing large data volumes. P12, for instance, suggests distinguishing between ``unclear'' and ``clearly threatening or criminally relevant'' posts, enabling a graded risk-ranking that could reduce workload when combined with manual review. Others propose using multiple classifiers to detect different forms of discriminatory content, reflecting organizational differences in categorization schemes \cite{baumler_harnessing_2025}.

Participants also highlight the value of information extraction, such as identifying names, dates, and locations, particularly for tracking mobilization efforts. Automated detection of calls for action, also in images, is seen as especially helpful, as relevant content can easily be missed across numerous groups -- ``a tool that tells you `here's a sharepic' would help a lot.'' (P8). However, limitations remain, for example in recognizing names in audio transcripts, a known challenge in current ASR systems \cite{ayache2024whispernerunifiedopennamed}. Other techniques such as facial recognition are considered problematic due to data protection regulation.
Beyond content-level analysis, participants express interest in user- and network-level insights, including tracking key actors, content origins, and forwarding patterns to reveal coordination. While such analyses are seen as valuable for understanding dissemination dynamics, they also raise ethical and legal concerns related to privacy and surveillance \cite{conway_online_2021,massanari_rethinking_2018}.
A smaller number of participants mention advanced analytical methods such as video summarization, topic modeling, and burst detection to support triage and early warning. Across these discussions, participants repeatedly emphasize both the potential and the limitations of AI-based analysis, underscoring the need for cautious human-AI collaboration -- a theme we return to in Section~\ref{ssec:human_ai}.

\subsubsection{Output and Documentation} 
\label{sssec:output}
Participants emphasize the need for automated saving of content and the generation of legally compliant documentation.
A central concern is the ability to capture content before it is deleted, as relevant posts, such as incitements to violence or invitations to illegal events are often removed after a certain time. Participants therefore stress that collection intervals should be configurable, ranging from minutes to longer periodic snapshots. P14 further suggests integrating report creation directly into the tool, allowing users to flag content and ``output a report at the end'' for a specific topic.
Proper digital evidence handling is another key requirement. Existing tools are described as inadequate -- ``you can't save videos with it at all'' (P5) -- prompting calls for solutions that reliably archive multimedia content with timestamps for potential court use. As P5 notes, this is critical ``because the evidence is captured when it’s first discovered.''
Finally, participants highlight the potential to streamline criminal complaint processing, which currently involves manually transferring information from spreadsheets into reporting forms, a process described as highly inefficient.

\subsubsection{Acceptance of New Tools}

Tool acceptance depends on usability, a clear reduction in workload, and compliance with data protection requirements, ideally supported through certification. These criteria align with prior work calling for user-friendly reporting technologies that meet data security and evidentiary standards \cite{baumler_harnessing_2025}.
Participants further stress technical requirements such as open source, local deployment, and reliable data storage, motivated by both security and cost concerns. However, these demands often conflict with state-of-the-art AI methods, particularly due to their computational demands and restrictive licensing. Overall, participants emphasize that non-technical factors -- organizational practices, legal requirements, and trust -- strongly shape tool adoption.

\subsection{Perspectives on the Role of AI and Human-AI Collaboration (RQ 3)}
\label{ssec:human_ai}

\begin{figure}
    \centering
\includegraphics[width=\linewidth]{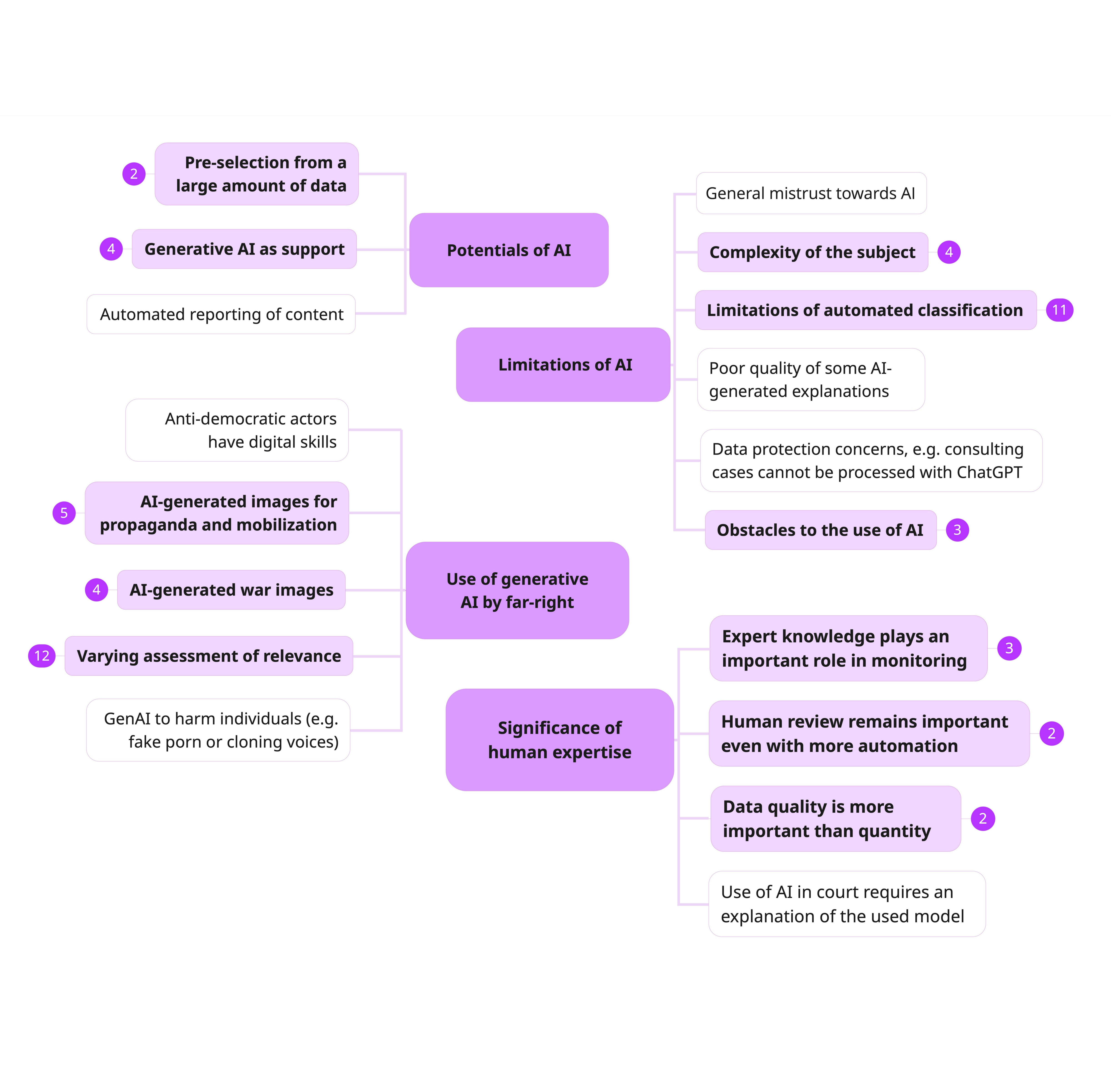}
    \caption{Thematic map section of CSO perspectives on the role of AI and human-AI collaboration in monitoring. Themes (purple nodes), subthemes (light purple nodes), and code counts (collapsed for space) are shown. White labels indicate individual codes not assigned to a subtheme.}
    \label{fig:human_ai}
    \Description{Thematic map section of CSO perspectives on the role of AI and human-AI collaboration in monitoring. Themes (purple nodes), subthemes (light purple nodes), and code counts (collapsed for space) are shown. White labels indicate individual codes not assigned to a subtheme.The thematic map was created based on qualitative coding of 15 interviews with CSO practitioners working for German-speaking organizations.} 
\end{figure}

During the interviews, participants discussed the role and limits of AI support and human-AI collaboration. The corresponding themes and subthemes are shown in Figure \ref{fig:human_ai}.

\subsubsection{Potentials of AI}
Despite some noting organizational mistrust toward AI, participants highlight its potential to support monitoring work by pre-filtering and prioritizing large volumes of data. As P4 notes, ``automation would help us really massively.'' Such uses align with prior work on claim detection and case prioritization \cite{hrckova_autonomation_2024,baumler_harnessing_2025}.
Participants also see promise in generative AI for assisting analysis and reporting, for example by identify far-right ideological traits in texts, or by summarizing insights across multiple sources, echoing findings from journalism and FC research \cite{liu_human-centered_2024,wolfe_implications_2024,cifliku_save_2025,procter_observations_2023}.

\subsubsection{Limitations of AI}
\label{sssec:limitations_AI_classification}
However, participants also question the use of AI, e.\,g. by citing misclassification risks that could produce a ``watered-down picture'' (P4) and expressing significant doubt that AI can adequately grasp the nuanced, context-dependent nature of harmful content in its varied expressions. These concerns mirror findings from OSINT, FC, and reporting CSOs \cite{juneja_human_2022,hrckova_autonomation_2024,mukhopadhyay_osint_2024,skopik_osint_nlp_2024,baumler_harnessing_2025}.
Further, existing models are not well suited to the linguistic and contextual demands of the domain. Participants note shortcomings for German-language content and phenomena such as antisemitism, which often ``works between the lines'' (P14). In CSCW terms, these concerns reflect challenges in heterogenous knowledge production regarding the translation of locally grounded concepts into standardized models \cite{star_institutional_1989,ackerman_sharing_2013}. P14 further highlights the substantial effort involved in training custom AI classifiers, alongside a lack of model sharing across CSOs, as ``those who train their own model often keep it to themselves due to the effort involved''.

\subsubsection{Use of Generative AI by the Far-Right}
Participants report mixed views on the relevance of generative AI used by far-right actors. Most believe it is primarily used for propagandistic material such as images or sharepics, while traditional disinformation strategies such as taking information out of context remain dominant. 
Others highlight the circulation of apparently AI-generated war imagery, which is seen as particularly concerning due to its potential to distort public perception.

\subsubsection{Significance of Human Expertise}
\label{sssec:human_expertise}
Across interviews, participants emphasize the central role of human expertise. AI is framed as a human-in-the-loop support tool rather than a replacement, aligning with established approaches to human-AI collaboration \cite{demartini_human_loop_2020}.
Some question whether AI is necessary at all: P15 emphasizes that human expertise is trusted by clients and perceived as more credible than AI-based classifications: ``We do a good job and I think it's easier for people to believe us.''
Others point to legal contexts where reliance on AI may complicate proceedings, as models would need to be explained in court. As P13 explains, ``human expertise is still considered more valuable than relying on AI,'' because if AI is involved ``you'd have to explain the model in court, which adds complexity''. These insights emphasize the role of trust in the adoption of new technologies, at organizational and at societal level. 

\subsection{Conceptual Monitoring Workflow}
\begin{figure}
    \centering
    \includegraphics[width=\linewidth]{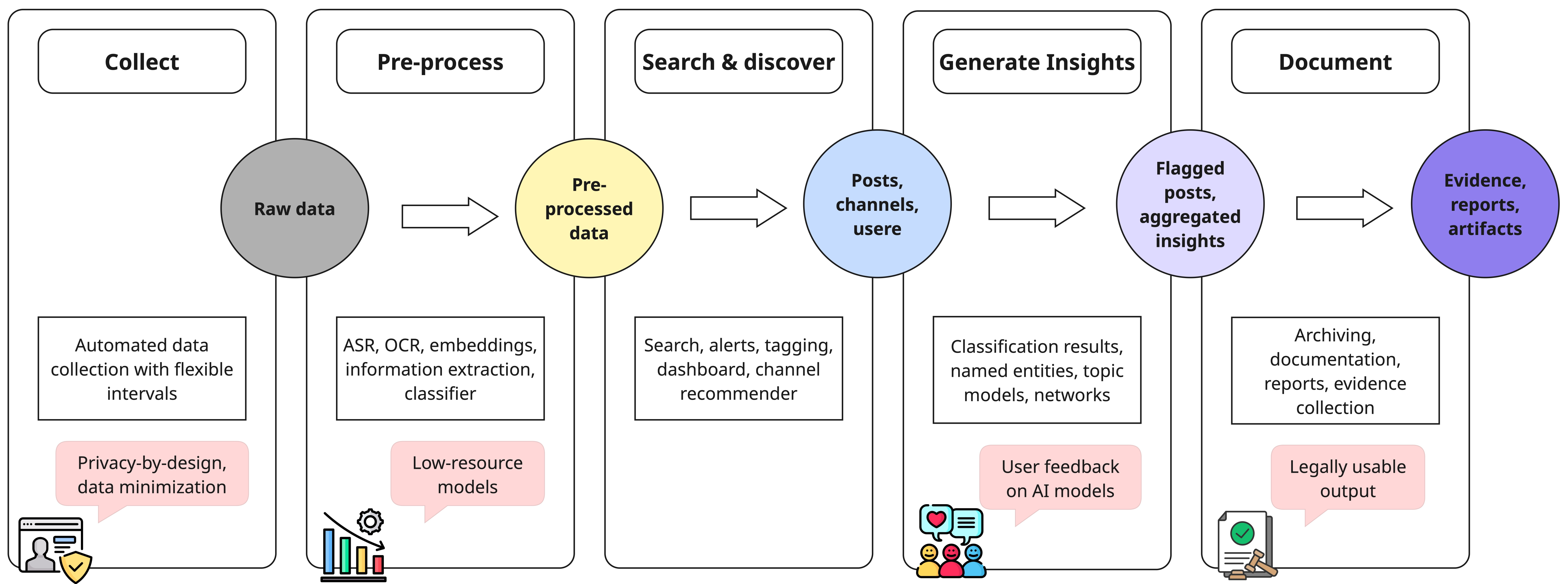}
    \caption{Conceptual overview of CSO monitoring workflow, aligned with technical, legal, and socio-technical requirements.}
    \label{fig:pipeline}
    \Description{A diagram showing the different steps of the derived pipeline alongside contextual aspects such as human-AI collaboration, legal and technical constraints}
\end{figure}

Based on our interview findings, we developed a conceptual monitoring workflow for CSO monitoring, shown in Figure~\ref{fig:pipeline}. This workflow synthesizes practitioners' described practices, constraints, and needs into a structured pipeline and instantiates the ``pipeline of practices'' discussed in prior work \cite{micallef_true_2022}. The resulting framework serves both as a guide for implementation of the prototype and as a bridge between technical development and real-world monitoring practices. Each stage of the workflow is aligned with specific design principles: 
\begin{itemize}
    \item \textbf{Data collection and storage} integrates \textit{privacy by design and legal usability} (\ref{sssec:legal_challenges}), ensuring that scraping modules capture metadata in a legally robust way while respecting data protection requirements \cite{baumler_harnessing_2025,hancock_seeing_2024,lakomy_open-source_2024}. 
   \item \textbf{Preprocessing and indexing} implement both \textit{multimodal data preprocessing} (\ref{sssec:challenge_multimodal}) and \textit{robust information extraction} (\ref{sssec:generate_insights}), making diverse content searchable and transforming noisy, unstructured data into structured formats \cite{cifliku_save_2025, juneja_human_2022,liu_human-centered_2024,ayache2024whispernerunifiedopennamed}. 
   \item \textbf{Search and retrieval} reflects the centrality of \textit{advanced and flexible search functionality} (\ref{sssec:enhanced_search}), enabling multilingual and multimodal content discovery through both lexical and semantic methods \cite{mukhopadhyay_osint_2024,beers_examining_2020,juneja_human_2022}. 
   \item \textbf{Dashboard and reporting} operationalize \textit{usability and workflow integration} (\ref{sssec:dashboard}), providing intuitive interfaces for tagging, saving searches, collaborative work, and generating reports \cite{cifliku_save_2025,baumler_harnessing_2025,liu_human-centered_2024}. 
   \item \textbf{Deployment and scalability} correspond to \textit{low-resource deployment and configurability} (\ref{sssec:data_access_handling}), ensuring that all modules are lightweight, open, and adaptable to diverse CSO infrastructures \cite{baumler_harnessing_2025,hummel_civil_2022,hulshof-schmidt_2024_2024,hancock_seeing_2024}.
\end{itemize}

Based on these insights, the monitoring prototype was extended in collaboration with the partner CSO, to include the functionalities shown in Table \ref{tab:functionalities}. 

\begin{table}[h]
\centering
\caption{Overview of the extended prototype's features.}
\label{tab:functionalities}
\begin{tabular}{p{3cm} p{10cm}}
\hline
\textbf{Component} & \textbf{Description} \\
\hline
Data collection &
Robust collection of historical and live Telegram messages with configurable time ranges and selectable attachment types. \\

Audio transcription &
Automatic transcription of audio messages using the open-source Whisper model (medium version by default; other model sizes configurable). \\

Lexical and semantic search &
Support for both keyword-based and embedding-based search across text and images, using Elasticsearch together with multilingual Sentence Transformer and CLIP models. \\

Channel recommendation &
Recommendation of related channels based on Telegram’s built-in ``Similar channels'' feature, used as an initial pragmatic default despite participant-reported limitations. \\

Automatic classification &
Fine-tuned BERT-based classifier for conspiracy-theory detection, deployable on CPU infrastructure, with integrated user feedback and data annotation support. \\

Interactive dashboard &
Web-based interface supporting saved searches, post-level statistics, manual content tagging, and collaborative use without advanced technical expertise. \\

Configurable system settings &
Environment-file configuration for data collection and retention policies, model selection (transcription, embeddings, classification), hardware utilization, and storage endpoints. \\

Modular containerized deployment &
Docker-based, modular architecture enabling flexible deployment and full system customization, with associated technical setup requirements. \\
\hline
\end{tabular}
\end{table}

\subsection{Findings from Workshops}
The workshops provided an opportunity to observe how the conceptual workflow and prototype features aligned with practitioners' monitoring practices. Participants extensively explored search functionality around their own thematic interests and emphasized the centrality of search for early-stage investigation. Features supporting the saving and organization of results -- such as tagging and saved searches -- were particularly valued. Several participants additionally requested export functions to support downstream analysis and reporting.

The sessions also surfaced organizational and infrastructural considerations. When multiple participants used the system simultaneously, the server reached its performance limits, revealing concrete bottlenecks and optimization needs. Beyond technical issues, participants repeatedly raised data protection concerns and exchanged organizational strategies for legally compliant monitoring, highlighting the importance of embedding legal and ethical considerations into tool design. 

Beyond tool testing, the workshops also functioned as sites of exchange between CSOs. Participants used the sessions to discuss investigative practices, constraints, and collaboration opportunities. 
We position these workshops as an initial co-design and formative engagement rather than a summative evaluation. They functioned as sites of mutual learning, where monitoring practices, constraints, and tool affordances could be collectively explored. In addition, a longer-term test with two civil society projects is currently underway to explore how the tool can be integrated into real-world monitoring work.

\section{Discussion and Recommendations}
Our findings highlight how the precarious operating conditions of CSOs -- limited resources, restricted platform access, and legal uncertainty -- systematically shape and constrain their engagement with technology. They lack direct platform access, tailored tools, and ongoing technical support, relying instead on labor-intensive, largely manual processes to identify and collect relevant content from the outside.

While automation is seen as potentially useful, particularly for expanding access to audio, image, and video data, participants express reservations about automated classification, which is often perceived as misaligned with CSO workflows. Instead, CSOs prioritize tools that support discovery, analysis and documentation across heterogeneous content formats. 
These findings foreground a gap between prevailing technological paradigms and the situated realities of civil society monitoring.

Across interviews, workflow conceptualization, and participatory workshops, we engaged this gap through an iterative process in which monitoring practices were translated into design concepts and a software prototype.

The following discussion synthesizes our empirical findings and derives design recommendations for CSO-oriented monitoring tools. The design recommendations aim at addressing the structural and infrastructural barriers that shape what technologies are realistically usable in CSO monitoring practice. Further, we outline stakeholder-specific recommendations for academic research and policymakers. Finally, we introduce the concept of a manual labor trap as an empirically grounded entry point for the development of a theory of collaborative sociotechnical scaling.

\subsection{Design Recommendations}
\label{ssec:design_recommendations}
Our findings highlight the need for monitoring tools that balance automation with human oversight, support multimodal content, ensure legal and ethical compliance, and remain realistically deployable in low-resource environments. Table \ref{tab:design_synthesis} synthesizes these insights by linking CSO practices and challenges with related work, core technical requirements, and resulting design recommendations.

An important challenge lies in the tension between privacy requirements and evidentiary needs. Monitoring tools must comply with data protection regulations, such as data minimization and secure deletion, while also preserving forensic-quality, legally admissible evidence. This requires careful trade-offs in data retention and export, and constraints the use of external services due to concerns around trust and compliance (rows 1, 2, 10 in Table \ref{tab:design_synthesis}). 
Fragmented and largely manual workflows further shape monitoring practice, limiting both intra- and inter-organizational collaboration. This underscores the need for integrated, user-friendly dashboards that support end-to-end processes, from data collection and analysis to report generation (row 3). Enabling effective collaboration additionally requires user management and  integration with existing workflows (row 4). 
The growing importance of multimodal and multilingual data constrasts with a limited availability of tools, highlighting the need to integrate ASR and OCR pipelines into search and analysis components (row 5). 
At the same time, the large volume of data encountered in online monitoring creates demand for automated filtering, ranking, and information extraction (row 8), and suggests the integration of automated content classification. However, the latter is met with notable skepticism due to concerns about reliability and transparency, reinforcing the importance of human-in-the-loop designs (row 6). 
Search emerges as a core functionality, requiring multimodal, and multilingual capabilities supported by both lexical and semantic methods (row 7).
Finally, resource and legal constraints strongly shape design decisions. Concerns related to privacy, security, cost, and trust in external services emphasize the importance of secure, cost-efficient local deployment models (rows 9, 10).

\begin{table*}[htp]
\small
\centering
\caption{Design synthesis mapping empirical findings to core technical needs and design recommendations.}
\label{tab:design_synthesis}
\begin{tabular}
{p{0.01\textwidth} p{0.18\textwidth} p{0.18\textwidth} p{0.18\textwidth} p{0.30\textwidth}}
\toprule

\textbf{} &
\textbf{Empirical\newline Finding} &
\textbf{Related Work} &
\textbf{Core Technical Need} &
\textbf{Design Recommendation} \\
\midrule

1 & Legal and ethical constraints of data collection and processing & Legal risks and constraints \cite{baumler_harnessing_2025,hancock_seeing_2024,lakomy_open-source_2024} &
Privacy-preserving data handling & Privacy-by-design architectures for data minimization and deletion \\ \hline

2 & Need for legally usable evidence & OSINT tools and digital evidence  \cite{mckeown_investigating_2014,cifliku_save_2025,baumler_harnessing_2025,skopik_osint_nlp_2024} &  Evidentiary robustness of collected content  & Forensic-quality capture with logged metadata (timestamps, URLs, IDs); export in standardized formats (e.\,g., PDF/A, WARC) suitable for legal review \\ \hline
 
3 & Fragmented, manual workflows & Monitoring as manual effort \cite{beers_examining_2020,das_state_2023,juneja_human_2022} &
Integrated, workflow-aware interfaces  &
Intuitive dashboards supporting content flagging, reporting \\ \hline

4 & Intra-organizational collaborative monitoring practices & CSOs as distinct actors, fragmented tooling \cite{juneja_human_2022,baumler_harnessing_2025,micallef_true_2022} &
User account management and seamless integration into existing CSO workflows  & User accounts to control for private and shared monitoring insights, shared labels and tags, data export functionalities into common formats (e.\,g. csv, PDF) \\ \hline

5 & Difficulty searching and analyzing multimodal content & Lack of multimodal tools \cite{cifliku_save_2025, juneja_human_2022,liu_human-centered_2024,ayache2024whispernerunifiedopennamed} &
Searchable representations of rich media content &
Integrated ASR and OCR pipelines to preprocess multimodal data
\\ \hline

6 & Skepticism toward automation & Algorithm aversion and human-in-the-loop \cite{juneja_human_2022,hrckova_autonomation_2024,mukhopadhyay_osint_2024,skopik_osint_nlp_2024,baumler_harnessing_2025} & Human-AI collaboration with transparency and controllability & Transparent communication of model limitations and errors to users (e.\,g., confidence cues, rationale surfaces, error reporting, feedback mechanisms) \\ \hline

7 & Centrality of search in monitoring work & Proactive and investigative work \cite{mukhopadhyay_osint_2024,beers_examining_2020,juneja_human_2022} & Advanced, flexible search across multilingual and multimodal data  & Robust lexical and semantic search across modalities using language-appropriate embedding models \\ \hline

8 & Large volumes of unstructured, noisy platform data & Filtering and ranking in FC and CM, information extraction in OSINT \cite{hrckova_autonomation_2024,baumler_harnessing_2025,mukhopadhyay_osint_2024} & 
Automated extraction of actionable information  &
Named Entity Recognition and information extraction models optimized for informal, real-world content (e.\,g., Telegram) \\ \hline

9 & Limited computational resources and technical infrastructure & Resource constraints \cite{baumler_harnessing_2025,hummel_civil_2022,hulshof-schmidt_2024_2024,noauthor_striving_2023,hancock_seeing_2024} &
Lightweight, adaptable, and maintainable systems &
Architectures with opinionated defaults; lightweight or distilled models deployable without GPU hardware \\ \hline

10 & Privacy, security, and cost concerns with external services & Legal constraints and CSO autonomy; trust and adoption concerns \cite{wolfe_implications_2024,cifliku_save_2025,hancock_seeing_2024,hrckova_autonomation_2024} & Secure and cost-efficient local deployment &
Minimized reliance on third-party APIs; prioritization of smaller, locally deployable models balancing performance, cost, and compliance \\
\bottomrule
\end{tabular}
\end{table*}

\subsection{Policy and Research Recommendations}
Our findings show that CSO monitoring is fundamentally sociotechnical and politically embedded, and cannot be addressed through tool design alone. We therefore outline empirically grounded policy and research recommendations in Table~\ref{tab:policy_research_implications} that complement our design recommendations. 

\begin{table}[h]
\centering
\small
\begin{tabular}{p{0.13\linewidth} p{0.85\linewidth}}
\toprule
\textbf{Theme} & \textbf{Recommendation} \\
\midrule
\multicolumn{2}{l}{\textbf{Policy}}\\
\addlinespace[2pt]
Legitimacy\newline \& role &
Formally recognize and mandate CSO online monitoring to strengthen legitimacy, trust, and inter-organizational collaboration.\\

Legal clarity &
Develop purpose-specific legal guidance for CSO monitoring and related activities (e.\,g., archiving, counseling), coordinated by national or federal data protection authorities, to reduce legal uncertainty and collaboration barriers.\\

Infrastructure \& funding &
Shift from short-term project funding to sustained public support for CSO infrastructure, collaboration, and open-source tool development, extending successful pilots into long-term operational funding.\\

\midrule
\multicolumn{2}{l}{\textbf{Research}}\\
\addlinespace[2pt]
Core pipelines &
Prioritize robust preprocessing pipelines (e.\,g., ASR, NER), especially for non-English and noisy real-world data, over (multimodal) classifiers, to enable scalable multimodal analysis and insight generation.\\

Discovery &
Support exploratory search and information discovery by studying CSO search practices and developing CSO-oriented tools (e.\,g., channel recommenders).\\

Domain\newline models &
Develop CSO-oriented, domain-specific models (e.\,g., for threats, antisemitism, racist mobilization) rather than generic toxicity or offensiveness classifiers.\\

Interfaces &
Investigate prompt-based LLM interfaces and flexible configurations to support adaptable, CSO-oriented workflows and non-expert querying.\\

Human-AI\newline collaboration &
Design lightweight, in-task human-in-the-loop feedback mechanisms to integrate practitioner expertise into iterative system improvement.\\

Inter-organizational collaboration &
Strengthen interdisciplinary and cross-sector research (e.\,g., law, political science, communication) and examine inter-CSO collaboration and networks.\\
\bottomrule
\end{tabular}
\caption{Policy and research recommendations for supporting CSO online monitoring.}
\label{tab:policy_research_implications}
\end{table}

On the policy side, a core challenge is structural misalignment: existing legal frameworks, funding models, and institutional arrangements do not reflect the long-term nature of CSO monitoring outside platform infrastructures or recognized professional workflows. We suggest three directions to address this. 
First, formal recognition of CSOs' societal role and clear institutional mandates could help to strengthen legitimacy, trust, and inter-organization collaboration, while enabling standardization, assurance of quality standards, and capacity-building. 
Second, legal uncertainty must be reduced through purpose-specific legal guidance for monitoring CSOs. This is particularly important given the risk of lawsuits or strategic information requests from antidemocratic actors, and the limited access to legal expertise among smaller organizations. Data protection authorities could play a coordinating role in producing accessible, sector-wide guidelines. 
Third, funding models should move beyond one-off project funding, which is structurally misaligned with the long-term nature of monitoring and coordination work. Governments and public institutions should instead invest in durable infrastructure -- including open-source tools co-developed with CSOs -- alongside sustained operational funding that enables gradual capacity-building.\footnote{Ironically, at the time of writing, Germany's main federal funding programme for civil society, \textit{Demokratie leben!}, is undergoing significant restructuring, leaving several democracy-focused CSOs in an existential funding crisis \cite{tagesschau_2026}. This development coincides with growing right-wing and far-right influence in Germany and across Europe -- a confluence of reduced civic support and democratic backsliding.}

Academic research should better recognize CSOs as underresearched stakeholders in digital governance and address the mismatch between current research priorities (e.\,g., hate speech classification) and practitioners' needs. 
To this end, greater emphasis should be placed on foundational capabilities rather than new task formulations, with increased investments in robust preprocessing pipelines (e.\,g., NER, ASR), especially for noisy, real-world, and non-English data. Further work is also needed to understand how CSOs search and explore unfamiliar spaces and discover emerging narratives. 

Given skepticism toward generic, English-centric AI models, research should support context-specific models tailored to CSO needs, such as detecting e.\,g., threats, antisemitism, or coordinated mobilization. This also includes exploring prompt-based LLM workflows through non-expert user studies.
In this context, we suggest that interactive systems with lightweight human feedback loops can enable continuous improvement and trust, provided they integrate seamlessly into everyday workflows and do not significantly increase the workload for CSOs.
Finally, as CSO challenges span legal, political, and technical domains, research should foster interdisciplinary collaborations including, e.\,g., organizational studies, law, communication, and political science to better understand hindrances to inter-organizational cooperation, and how CSOs navigate adversarial environments.

\subsection{The Manual Labor Trap and Inter-organizational Collaboration}

Our findings show that monitoring CSOs operate under precarious infrastructural conditions marked by limited funding, restricted platform access, and legal uncertainty. Participants consistently described a sociotechnical gap between what they aim to accomplish -- long-term, collaborative, and thematically diverse monitoring -- and what they can realistically support with limited IT capacity, unstable resources, and a lack of sustained infrastructure. Monitoring practices center on simple tools and ad-hoc workflows that are difficult to extend or maintain. From a CSCW perspective, CSOs rely on a fragmented installed base \cite{star_steps_1994,neumann_making_1996} held together through ongoing articulation work \cite{schmidt_taking_2011}. Efforts to extend this base often take the form of bricolage \cite{buscher_landscapes_2001,karanasios_digital_2022}, creatively recombining available tools to overcome local constraints. Under the described adversarial and resource-scarce conditions, however, bricolage hardens into what we conceptualize as a \textbf{manual labor trap}: labor-intensive and brittle arrangements consume the very resources needed to build more durable infrastructures. Over time, this self-reinforcing loop stabilizes low-capacity sociotechnical regimes increasingly misaligned with the scale and dynamism of far-right online activity. Being caught in this trap also prolongs exposure of individuals conducting monitoring to harmful content, with likely long-term impacts on psychological wellbeing, e.\,g., by increasing fear, mistrust, or resignation among individuals. Scaling infrastructures is therefore not only a matter of efficiency, but also of protection.

Importantly, the limitations producing the manual labor trap are not primarily technical; financial and legal constraints fundamentally shape which infrastructural choices are available. Under these conditions, purely individual approaches to technological development leave little room for meaningful scaling. Chronic underfunding, legal uncertainty, restricted platform access, and ongoing maintenance demands make it unlikely that single CSOs can independently build and sustain the infrastructures their work increasingly requires. In this sense, the manual labor trap appears not as a temporary phase but a structurally reproduced outcome of isolated operation. From this perspective, inter-organizational collaboration becomes a plausible horizon for sociotechnical scaling.

To synthesize these dynamics, we propose the development of a theory of \textbf{collaborative sociotechnical scaling} in CSO monitoring -- a theory that situates CSO infrastructures within their structural and technological constraints, contrasting the manual labour trap with collaborative arrangements that could support the development and maintenance of shared infrastructures.

Respective theorizing could also help to address the tension between demands for flexible and configurable infrastructure that account for diverse and evolving monitoring interests, and usability requirements of non-technical users: Under collaborative sociotechnical scaling, CSOs with limited technical expertise could benefit from shared knowledge, pooled development capacity, and the longer-term creation of user-friendly configuration interfaces. 

The concept of the manual labor trap offers an entry point for theorizing sociotechnical scaling in civil society monitoring. However, collaboration is not a neutral solution: pooling resources raises questions of governance, ownership, and power, and risks reproducing asymmetries between better- and less-resourced actors. A fuller account of collaborative sociotechnical scaling would need to treat collaboration as a contested transformation shaped by economic, legal, and political conditions and uneven agency.
Developing such an account calls for methodological approaches beyond tool-focused design research. Ethnographic, organizational, or even historical studies of existing or past collaborative settings -- examining when and why they succeed or fail -- would help surface the conditions under which shared infrastructures become sustainable. Trust, competition for resources, and the unequal capacity to contribute to and benefit from collaboration are likely central dynamics. Tech collectives that work across multiple CSOs as intermediary hubs, for example the CSO Tactical Tech,\footnote{https://tacticaltech.org/} represent a promising site for such inquiry, as they concentrate collaboration processes and make their tensions visible. Historical accounts of sociotechnical scaling in analogous civil society contexts could further illuminate the structural and political conditions that enable or foreclose such transitions. Throughout, attention to invisible labor and articulation work \cite{star_layers_1999,schmidt_taking_2011} will be essential: collaborative infrastructures do not maintain themselves, and the work of sustaining them is easily obscured, inequitably distributed, and rarely funded. We take this as an important direction for future CSCW work.

\section{Limitations}

Our study focuses on German-speaking organizations. Although diverse, they represent a limited sample of CSOs engaged in monitoring across Europe. The transferability of our findings thus requires caution. Constraints such as limited data access and scarce resources likely extend across political and socio-economic contexts -- and may be more severe in lower-income settings where limited connectivity and technological infrastructure restrict access to computational and AI-based tools. On the other hand, legal, political, and socio-economic conditions vary widely, shaping different levels of vulnerability. Our participating CSOs operate in a Western, European democracy; organizations elsewhere may face additional risks, including state repression \cite{greig_civil_soc_2024}. In more authoritarian contexts, the threat model shifts fundamentally: anonymity and operational security become central requirements, and the value of legally robust evidence for domestic proceedings may be lower where such processes are inaccessible or unsafe, with international accountability mechanisms mattering more instead. In this context, transnational collaboration -- with external CSOs absorbing public-facing exposure while threatened organizations focus on evidence collection -- may itself constitute a form of sociotechnical scaling, distributing both risk and capacity across organizational (and national) boundaries.
The role of online platforms also differs by political and regulatory context, over time, and according to the thematic focus of CSOs. Even within our sample, practices and needs varied, underscoring the difficulty of developing design implications that are both generalizable and flexible. Core needs such as multimodal data processing, robust search, and legal evidence handling appear broadly relevant, but must be adapted to organization-specific conditions. Future research should include a broader range of European and international CSOs to test and refine our findings.

The interviews were conducted with monitoring professionals rather than AI or software experts. Some participants struggled to articulate technical needs beyond existing practices, potentially narrowing perspectives. In some cases, the interviewer suggested potential features, prompting participants to consider aspects they may not have raised themselves, as clarified in our methodology. 

While qualitative interviews provide a solid foundation as our main method, a more in-depth requirements analysis would benefit from additional methods, such as user stories and use cases, personas, contextual inquiries, design probes, and requirement prioritization techniques. To begin to address this limitation, we have incorporated workshops and are conducting longitudinal testing with two organizations.

\section{Conclusion}
Our study advances the understanding of CSOs as crucial yet understudied actors in the development and application of monitoring technologies. The precarious situation of CSOs, their diverse activities, and the multiplicity of obstacles they face underscore the heterogeneity of the field. This heterogeneity requires technological solutions that are flexible and adaptable, capable of accommodating varied organizational practices and thematic foci. Our findings further highlight that while CSOs recognize the potential of AI -- particularly in expanding access to otherwise inaccessible information embedded in audio, image, and video content -- there is widespread skepticism about automated classification systems. Such systems are perceived as misaligned with CSO workflows and especially problematic in legal and evidentiary contexts, where transparency, accountability, and human credibility are indispensable. 

The contributions of our work are threefold: first, we provide an in-depth, empirically grounded account of civil society monitoring as a form of collaborative, infrastructural, and epistemic work conducted under precarious, adversarial, and legally uncertain conditions. In doing so, we extend existing understandings in CSCW of how sociotechnical systems are maintained and made workable outside well-resourced institutional settings. 
Second, we translate these insights into actionable design and research implications. We articulate design considerations for flexible monitoring infrastructures focussing on search, (multimodal) data preprocessing, evidence collection and documentation, that remain adaptable across thematic domains. Third, we introduce the manual labor trap as an empirically grounded concept that explains why monitoring CSOs tend to remain locked into labor-intensive, low-capacity arrangements, and propose to develop a CSCW-grounded theory of collaborative sociotechnical scaling to better understand how alternative, collaborative modes of organization could reconfigure these dynamics. 

\received{May 13, 2025}
\received[revised]{January 13, 2026}
\received[accepted]{April 9, 2026}

\bibliographystyle{ACM-Reference-Format}
\bibliography{main}

\appendix

\section{Appendix}

\subsection{Participants}

\begin{table}[h!]
        \caption{Overview of interview participants, including the thematic focus of the employing organization (O), the participant's (P) main activity, and whether they have formal technical education.}
    \label{tab:participants_detail}
    \centering
    \small
    \begin{tabular}{|l|l|l|l|l|}
    \hline
       \textbf{Id} & \textbf{Thematic Focus (O)} & \textbf{Main Activity (P)} & \textbf{Technical Education (P)} \\ \hline
        P1 & Far-right & Consulting & No \\ \hline
        P2 & Antisemitism & Monitoring & No \\ \hline
        P3 & Far-right, conspiracy theories & Archiving, education & No \\ \hline
        P4 & Various phenomena & Consulting & No \\ \hline
        P5 & Various phenomena & Consulting & No \\ \hline
        P6 & Far-right & Monitoring & No \\ \hline
        P7 & Far-right & Monitoring & No \\ \hline
        P8 & Far-right & Archiving, analysis & No \\ \hline
        P9 & Far-right & Consulting & No \\ \hline
        P10 & Far-right & Monitoring  & No \\ \hline
        P11 & Conspiracy theories & Monitoring & No \\ \hline
        P12 & Various phenomena & Monitoring, analysis & Yes \\ \hline
        P13 & Human rights violations & Software development & Yes \\ \hline
        P14 & Antisemitism, racism, conspiracy theories & Monitoring & Yes \\ \hline
        P15 & Far-right & Consulting & No \\ \hline
    \end{tabular}
\end{table}

\subsection{Interview Guidelines}
\label{ssec:interview_guidelines}

\subsubsection{Information Phase}
\begin{itemize}

    \item Thanks for the willingness to be interviewed.
    \item Information about research + its objectives.
    \item Information about confidentiality.
\end{itemize}

\subsubsection{Warm-up Phase}
Have the work of the organisation / person explained in general terms. 
\begin{itemize}

    \item What can you tell me about your work / the work of your organisation? 
    \item What are the focal points (main topics) of your organisation?
    \item What is the main focus of your work? (e.\,g. counselling, monitoring, educational work)
\end{itemize}

\subsubsection{Main Phase}

\paragraph{General analysis / research}

\begin{itemize}
    \item What role does analysing anti-democratic actors online play in your work? 
    \item Which phenomena or categories are particularly relevant? 
    \item Which social media or services are particularly relevant for you?
    \item What role does Telegram play for you? 
    \item Which data/media types are of particular interest to you and why?  (text, image, audio, video)
    \item Which methods and processes do you use?
    \item Which languages are relevant?
\end{itemize}

\paragraph{Current workflow}
\begin{itemize}
    \item How do you currently go about analysing or researching -- could you describe a typical process as specifically as possible?
    \item What challenges or problems do you encounter? (e.\,g. technical or legal)
    \item What human and/or financial resources are available for monitoring/analysis in your organisation?
\end{itemize}

\paragraph{Existing services}
\begin{itemize}
    \item Do you already have software that you use for research/analysis? 
    \item What exactly do you find helpful about it? 
    \item Is there anything in it that you think is missing or that you would like to see better/differently?
\end{itemize}

\paragraph{AI}
\begin{itemize}
    \item  Where could AI best support you in your work? 
    \item What would be the ideal functionality for you?
    \item What about the use of Generative AI? 
    \item What role does generated content play in your monitoring work?
\end{itemize}

\paragraph{Other}
\begin{itemize}
    \item  How would you describe your technical infrastructure?
    \item Do you exchange information with others to keep each other up to date? 
    \item Do you share data with others? 
\end{itemize}

\subsubsection{Closing Phase}
\begin{itemize}
    \item Is there anything else you would like to add? 
    \item Information about next planned steps (publication of results, workshops to test the software)
    \item Thank participants.
\end{itemize}

\end{document}